\documentclass[showkeys,twocolumn]{revtex4}
\begin{document}
\title{\huge\textbf{
Second and first order phase transition in three dimension Gross-Neveu
model}}\thanks{This work was partially supported by National Natural Science
Foundation of China.}
\author{\Large Zhou Bang-Rong}
\affiliation{Department of Physics, Graduate School of the Chinese Academy of Sciences,  Beijing 100039}
\begin{abstract}
Symmetry restoring phase transitions in three dimension Gross-Neveu model are shown
to be second order at finite temperature $T$ and first order at $T=0$ and finite
chemical potential $\mu$ by critical analysis of the dynamical fermion mass based
on the gap equation. The latter is further verified by effective potential analysis.
The resulting tricritical point is $(T,\mu)=(0,m(0))$, where $m(0)$ is the dynamical
fermion mass at $T=\mu=0$. Physical difference between the above second and first
order phase transition is illustrated by means of variations of thermodynamical
particle density.
\end{abstract}
\maketitle
The Gross-Neveu (GN) model \cite{kn:1} with four-fermion interactions is a good laboratory
to research symmetry restoring phase transitions at high temperature and high density
\cite{kn:2,kn:3,kn:4,kn:5}. One believes that some phase structure of Quantum
Chromodynamics (QCD), e.g. a tricritical point in temperature $T$-chemical potential
$\mu$ phase diagram, i.e. the point where second order and first order phase
transition meets, could appear in such type of models \cite{kn:6}. Existence of a
tricritical point is of both theoretical and experimental significance. Since the
three dimension (3D) GN model is easy to deal with analytically, and in addition,
in this model there is still a question to be clarified further of that whether first
order phase transition exists or not in it \cite{kn:5,kn:6}, it is interesting to
make a careful analysis of the phase transition of this model at finite temperature,
especially at high density. In this paper, we will report our results and indicate
that a tricritical point exists indeed in 3D  GN model. Our method is to analyze
critical behaviors of the dynamical fermion mass as the order parameter of symmetry
breaking based on the gap equation obeyed by the dynamical fermion mass and on the
corresponding effective potential at finite temperature and chemical potential. The
discussions will be made in the fermion bubble diagram approximation which is
equivalent to the leading order of $1/N$ expansion. \\
\indent We first briefly review the used model \cite{kn:7}. The Lagrangian will be
written
by
\begin{equation}
{\cal L}(x)=\sum_{k=1}^N\bar{\psi}^k(x)i\gamma^{\mu}\partial_{\mu}\psi_k(x)+
              \frac{g}{2}\sum_{k=1}^N[\bar{\psi}^k(x)\psi_k(x)]^2,
\end{equation}
where $\psi_k(x)$ is a spinor with $N$ "color" components. In 3 dimensions, the
coupling constant $g$ has mass dimension -1 thus the model is perturbatively
non-renormalizable however it is renormalizable in $1/N$ expansion \cite{kn:4}. We
take the $\gamma^{\mu}\ (\mu =0,1,2)$ to be $2\times 2$ matrices and relate them to
the Pauli  matrices $\sigma^i (i=1,2,3)$ by $\gamma^0=\sigma^3$,
$\gamma^1=i\sigma^1$ and $\gamma^2=i\sigma^2$. Noting that in three dimensions there
is no "$\gamma_5$" matrix which anticommutes with all the three $\gamma^{\mu}$
matrices. Consequently ${\cal L}(x)$ has no chiral symmetry. However, the action of
the system is invariant under the special ${\cal P}_1$ and ${\cal P}_2$ parity
transformation (omitting the index $k$ of $\psi$)
\[
\psi(t, x^1, x^2)\stackrel{{\cal P}_1}{\longrightarrow}
\gamma^1\psi(t,-x^1, x^2),
\]
\[
\psi(t, x^1, x^2)\stackrel{{\cal P}_2}{\longrightarrow}
\gamma^2\psi(t,x^1, -x^2)
\]
and the time reversal
\[
\psi(t, x^1, x^2)\stackrel{{\cal T}}{\longrightarrow}
\gamma^2\psi(-t,x^1, x^2).
\]
If the four-fermion interactions in Eq.(1) will lead to the dynamical mass term
$-m\bar{\psi}\psi$, then the discrete symmetries ${\cal P}_1$, ${\cal P}_2$ and
 ${\cal T}$ will be spontaneously broken. These broken symmetries, however, could
be restored at high temperature and/or high density. This is just the problem we will
discussed.\\
\indent At zero-temperature $T=0$ and zero chemical potential $\mu=0$, assume the
four-fermion interactions $g\sum_{k=1}^N{(\bar{\psi}^k\psi_k)}^2/2$ can lead to the
fermion condensates $\sum_{k=1}^N\langle\bar{\psi}^k\psi_k\rangle\neq 0$, then we
will obtain the gap equation obeyed by the dynamical fermion mass $m(0)$
\begin{equation}
1=2gN\int\frac{id^3l}{{(2\pi)}^3}\frac{1}{l^2-m^2(0)+i\varepsilon}
\end{equation}
After the Wick rotation, angular integration and introduction of $3D$ Euclidean
momentum cut-off $\Lambda$, Eq.(2) becomes
\begin{equation}
1=\frac{gN\Lambda}{\pi^2}\left[1-
       \frac{m(0)}{\Lambda}\arctan\frac{\Lambda}{m(0)}\right].
\end{equation}
For sufficiently strong coupling $gN$ that $gN\Lambda/\pi^2>1$, Eq.(3) can be
satisfied. When $T\neq 0$, we must replace the vacuum expectation value
$\sum_{k=1}^N\langle\bar{\psi}^k\psi_k\rangle$ by the thermal expectation value
$\sum_{k=1}^N\langle\bar{\psi}^k\psi_k\rangle_T$. In the real-time formalism of
thermal field theory, this means the substitution of the fermion propagator \cite{kn:8}
\begin{eqnarray}
\frac{i}{\not{l}-m(0)+i\varepsilon}&\rightarrow &
  \frac{i}{\not{l}-m+i\varepsilon}-2\pi\delta(l^2-m^2)\nonumber \\
 && \times(\not{l}+m)\sin^2\theta(l^0,\mu)
\end{eqnarray}
\noindent with
\begin{equation}
\sin^2\theta(l^0, \mu)=\frac{\theta(l^0)}{\exp[\beta(l^0-\mu)]+1}
                          +\left(\matrix{l^0\to -l^0\cr
                                         \mu\to -\mu\cr}\right)
\end{equation}
where $\beta=1/T$ and $m\equiv m(T, \mu)$ is the dynamical fermion mass at finite
temperature $T$ and finite chemical potential $\mu$.  As a result, the gap equation
at $T\neq 0$ becomes
\begin{equation}
1=\frac{gN\Lambda}{\pi^2}\left[
   1-\frac{m}{\Lambda}\arctan\frac{\Lambda}{m}-\frac{\pi}{2\Lambda}F_2(T,\mu,m)
   \right],
\end{equation}
where
\begin{equation}
F_2(T,\mu,m)=T\ln\left[1+e^{-(m-\mu)/T}\right]+(-\mu\to \mu).
\end{equation}
Considering the gap equation (3) at $T=0$ and the fact that the momentum cut-off
$\Lambda$ may be high enough so that
$\arctan\frac{\Lambda}{m(0)}=\arctan\frac{\Lambda}{m}=\frac{\pi}{2}$, we will
reduce Eq.(6) to
\begin{equation}
m(0)= m+F_2(T,\mu,m).
\end{equation}
It is easy to verify that $\partial F_2(T,\mu,m)/\partial T>0$ and $\partial
F_2(T,\mu,m)/\partial \mu>0$, i.e. $F_2(T,\mu,m)$ is respectively a monotone
increasing function of temperature $T$ and chemical potential $\mu$. As a result,
it can be seen from Eq.(8) that, as $T$ and/or $\mu$ increase, $m$ must decrease and
finally go to zero at a critical point $(T_c,\mu_c)$. The critical equation to
determine the point $(T_c,\mu_c)$ will be
\begin{eqnarray}
m(0)&=&F_2(T_c,\mu_c,m=0) \nonumber \\
    &=&T_c\left[\ln (1+e^{\mu_c/T_c})+(\mu_c\to -\mu_c)\right].
\end{eqnarray}
In the above discussions, we have actually used an implicit assumption, i.e. $m$ could
go to zero continuously as $T$ and/or $\mu$ increase. This just represents the
characteristic of a second order phase transition. Hence the $T-\mu$ phase diagram
given by Eq.(9) will be the one of second order phase transition. The critical behavior
of $m$ near a critical point can be derived from Eqs.(8) and (9). Since near a critical
point $(T_c,\mu_c)$, $m\approx 0$, thus $m/T\ll 1$ for a finite temperature $T$. Then
Eq.(8) with Eq. (7) will lead to
\begin{widetext}
\begin{equation}
m(0)=T\ln\left(1+e^{\mu/T}\right)+T\ln\left(1+e^{-\mu/T}\right)+
      \frac{m^2}{2T[1+\cosh(\mu/T)]}+{\cal O}\left(\frac{m^3}{T^3}\right).
\end{equation}
Substituting Eq.(9) with $\mu_c$ replaced by $\mu$ into Eq.(10) we will  obtain the
critical behavior of $m^2$ near $T_c$ when $\mu$ is fixed
\begin{equation}
m^2=2T\left(1+\cosh \frac{\mu}{T}\right)\left\{
      \ln \left[2\left(1+\cosh \frac{\mu}{T_c}\right)\right]-
      \frac{\mu}{T_c}\tanh \frac{\mu}{2T_c}\right\}(T_c-T) \ \
      {\rm when} \ T\alt T_c
\end{equation}
\end{widetext}
Substituting Eq.(9) with $T_c$ replaced by $T$ into Eq.(10) we will  obtain the
critical behavior of $m^2$ near $\mu_c$ when $T$ is fixed
\begin{eqnarray}
&m^2&=2T\sinh(\mu/T)(\mu_c-\mu), \nonumber \\
&&{\rm when} \ \mu\alt \mu_c \ {\rm and} \ T\neq 0.
\end{eqnarray}
Eqs. (11) and (12) show that when $T\neq 0$, the symmetry restoring phase transitions
whether at $T_c$ or at $\mu_c$ are second order.  However, the above analyses are
inapplicable to the case of $T=0$, since the approximate expansion (10) is obviously
not valid in this case. \\
\indent For $T=0$, we must come back to the gap equation (8). Considering that
\begin{equation}
\lim \limits_{T\to 0}F_2(T,\mu,m)=\left\{\matrix{0 &{\rm when}&\mu\leq m\cr
                                                \mu-m&{\rm when}&\mu>m\cr}\right.,
\end{equation}
we can obtain from the $T\to 0$ limit of Eq.(8) that
\begin{eqnarray}
m(0)&=&m, \ {\rm when} \ \mu\leq m \\
m(0)&=&\mu, \ {\rm when} \ \mu>m
\end{eqnarray}
Eq.(14) can be changed into that $m=m(0)$, when $\mu\leq m=m(0)$. Then Eq.(15) will
become $m(0)=\mu$, when $\mu>m(0)$, two contradictory formulas each other. This
implies that the gap equation can not be satisfied when $\mu>m(0)$ thus there is no
symmetry breaking and the order parameter $m$ must be zero. The above result can be
summarized as
\begin{equation}
m=\left\{\matrix{m(0) &{\rm when} &\mu\leq m(0)\cr
0 &{\rm when} &\mu>m(0)\cr}\right..
\end{equation}
Eq.(16) indicates that when $T\to 0$, the order parameter $m$ will jump from $m(0)$
down to 0 as $\mu$ crosses over the critical chemical potential $\mu_c=m(0)$, hence
the phase transition is first order. This conclusion can also be verified by an
effective potential analysis. It can be proven that \cite{kn:9} the gap equation (8) may also come from the extreme value condition of the effective potential
$ V_{eff}^{(3)}(T,\mu,m)$, i.e. $\partial V_{eff}^{(3)}(T,\mu,m)/\partial m=0$,
where
\begin{equation}
\frac{\partial V_{eff}^{(3)}(T,\mu,m)}{\partial m}=
\frac{m}{2\pi}\left[m-m(0)+F_2(T,\mu,m)\right],
\end{equation}
Thus we can obtain from Eq.(17) by integration
\begin{eqnarray}
V_{eff}^{(3)}(T,\mu,m)&=&\frac{1}{2\pi}\int_0^mdm'm'\left[m'-m(0)\right.
\nonumber \\
&&\left.+F_2(T,\mu,m')\right]
\end{eqnarray}
which has been normalized to $V_{eff}^{(3)}(T,\mu,m=0)=0$. By means of Eq.(13) we
may obtain the $T\to 0$ limit of $V_{eff}^{(3)}(T,\mu,m)$
\begin{eqnarray}
V_{eff}^{(3)}(T=0,\mu,m)=\frac{1}{12\pi}
\left\{3m^2[\mu\theta(\mu-m)\right.&&\nonumber \\
\left.-m(0)]+\theta(m-\mu)(2m^3+\mu^3)\right\}.&& \nonumber \\
\end{eqnarray}
The shape of $V_{eff}^{(3)}(T=0,\mu,m)$ will vary as $\mu$ increases.\\
1) $\mu=0$. $V_{eff}^{(3)}(T=0,\mu,m)$ will have the maximum point $m=0$ and the
minimum point $m=m(0)$ with $V_{eff}^{(3)}[T=0,\mu,m(0)]=-\frac{1}{12\pi}m^3(0)$.
This simply represents the spontaneous symmetry breaking at $T=\mu=0$.\\
2) $\mu<m(0)$. $V_{eff}^{(3)}(T=0,\mu,m)$ has the same maximum point $m=0$ and the
same minimum point $m=m(0)$, but the minimum
$V_{eff}^{(3)}[T=0,\mu,m=m(0)]=\frac{1}{12\pi}[\mu^3-m^3(0)]$ will rise as $\mu$
increases. \\
3) $\mu=m(0)$. The effective potential now becomes
\begin{eqnarray*}
&&V_{eff}^{(3)}[T=0,\mu=m(0),m]\\
&&=\frac{1}{12\pi}\theta[m-m(0)]\left[m^3(0)-3m^2m(0)+2m^3\right]
\end{eqnarray*}
which indicates that the total real axis segment $0\leq m\leq m(0)$ including the
original $m=m(0)$ when $\mu<m(0)$ will be the minimum points of
$V_{eff}^{(3)}[T=0,\mu=m(0),m]$.\\
4) $\mu>m(0)$. $V_{eff}^{(3)}[T=0,\mu>m(0),m]$ will only have a minimum point $m=0$
and this implies that the broken symmetries will be restored. Obviously, the critical
chemical potential $\mu_c$ should be  determined by the condition
\begin{equation}
\left.\frac{\partial^2 V_{eff}^{(3)}(T=0,\mu,m)}{\partial m^2}\right|_{m=0}
=\mu-m(0)=0,
\end{equation}
i.e. at $\mu_c$, $m=0$ must change from a maximum point into a minimum point.
Eq.(20) gives that $\mu_c=m(0)$. As $\mu$ increases and crosses over $m(0)$, the
minimum point $m$ of $V_{eff}^{(3)}(T=0,\mu,m)$ will jump from $m(0)$ down to 0.
This will reproduce Eq.(16) and verify that $(T,\mu)=(0,m(0))$ is a first order phase
transition point. On the other hand, it is noted that  $(T,\mu)=(0,m(0))$ is also
a solution of the critical equation (9) corresponding to second order phase transition
when we take $\mu_c=\mu$ and the limit $T_c=T \to 0$. Hence we can reasonably conclude
that  $(T,\mu)=(0,m(0))$  will be a tricritical point in the $T-\mu$ phase diagram
of $3D$ GN model. \\
\indent The physical difference between the above second order and first order phase
transition can be clearly displayed by variations of thermodynamical particle density
near critical points. We will respectively consider the phase transitions at $T\neq
0$ and $T=0$. The general total density of fermions with mass $m$ in 3 dimensions
may be expressed by
\begin{widetext}
\begin{eqnarray}
n(T,\mu,m)&=&\int\frac{d^2\stackrel{\rightharpoonup}{p}}{(2\pi)^2}\left[
\frac{1}{e^{\beta(\sqrt{{\stackrel{\rightharpoonup}{p}}^2+m^2}-\mu)}+1}-(-\mu\to \mu)\right]\nonumber \\
&=&\frac{m^2}{2\pi}\int_1^{\infty}dz \ z\left[\frac{1}{e^{yz-r}+1}-(-r\to r)\right], \ y=\beta m,\ r=\beta\mu, \nonumber \\
&=&\frac{Tm}{2\pi}\ln\frac{1+e^{-y+r}}{1+e^{-y-r}}+\frac{T^2}{\pi}\sum_{k=1}^{\infty}\frac{(-1)^{k+1}}{k^2}\sinh(kr)e^{-ky}.
\end{eqnarray}
For the model of dynamical spontaneous symmetry breaking, $m$ should be identified
with the dynamical fermion mass. When $T\neq 0$, the phase transition is second order
thus near a critical point $(T_c,\mu_c)$, $y\ll 1$, we can make the expansion of
$n(T,\mu,m)$ in the powers of $y$ up to ${\cal O}(y^2)$ order and obtain
\begin{equation}
n(T,\mu,m)\simeq n(T,\mu,m=0)-\frac{m^2}{4\pi}\frac{e^r-1}{e^r+1}, \ {\rm when} \ \frac{m}{T}\ll 1.
\end{equation}
Now if we consider the phase transition at finite chemical potential and $T\neq 0$,
then can get by using Eq.(12)
\begin{equation}
n(T,\mu,m)=\left\{\matrix{ n(T,\mu,m=0)-\frac{T(e^r-1)}{2\pi(e^r+1)}\left(\sinh\frac{\mu}{T}\right)(\mu_c-\mu), &{\rm when} \ \mu\alt \mu_c \cr
n(T,\mu,m=0),                    &{\rm when} \ \mu > \mu_c \cr}\right..
\end{equation}
However,
\begin{equation}
\frac{\partial n(T,\mu,m)}{\partial \mu}\simeq
\left\{\matrix{ \frac{\partial n(T,\mu,m=0)}{\partial \mu}
+\frac{T(e^r-1)}{2\pi(e^r+1)}\sinh\frac{\mu}{T}, &{\rm when} \ \mu\alt \mu_c \cr
\frac{\partial n(T,\mu,m=0)}{\partial \mu},  &{\rm when} \ \mu > \mu_c \cr}\right..
\end{equation}
\end{widetext}
Eqs.(23) and (24) indicate that when $\mu\to \mu_c$, $n(T,\mu,m)$ will increase
continuously up to $n(T,\mu,m=0)$ but $\partial n(T,\mu,m)/\partial \mu $ will have
a discontinuous jump across over $\mu_c$. These just show second order feature of
the phase transition. For discussion of the phase transition at $T=0$, we first find
out the $T\to 0$ limit of $n(T,\mu,m)$ from Eq.(21)
\begin{equation}
n(T=0,\mu,m)=\theta(\mu-m)\frac{\mu^2-m^2}{4\pi}.
\end{equation}
Since the phase transition at $T=0$ and $\mu=\mu_c=m(0)$ is first order, we can
substitute Eq.(16) into Eq.(25) and obtain
\begin{equation}
n(T=0,\mu,m)=\left\{\matrix{0, &{\rm when} \ \mu\leq m(0) \cr
\frac{\mu^2}{4\pi}, &{\rm when} \ \mu >m(0) \cr}\right..
\end{equation}
Eq.(26) shows that when $T=0$, as $\mu$ crosses over the critical point $\mu_c=m(0)$,
the particle density will jump from 0 to $\mu^2/4\pi$ (the density of massless
fermions) and this just represents a characteristic of first order phase
transition.\\

\end{document}